# Conductance Fluctuations in PbTe Wide Parabolic Quantum Wells

J. Oswald[1], G. Heigl[1], G. Span[1], A. Homer[1], P. Ganitzer[1], D.K. Maude[2], J.C. Portal[2]

[1]Institute of Physics, Montanuniversiaet Leoben, Franz Josef Str. 18, A-8700 Leoben, Austria

[2]High Magnetic Field Laboratory, CNRS, 25 Avenue des Martyrs, BP 166 Grenoble, France

*Abstract*

*We report on conductance fluctuations which are observed in local and non-local magnetotransport experiments. Although the Hall bar samples are of macroscopic size, the amplitude of the fluctuations from the local measurements is close to $e^2/h$. It is shown that the fluctuations have to be attributed to edge channel effects.*

## 1. Introduction

Most of the experiments which agree well with the theory of universal conductance fluctuations (UCF) [1,2] have been performed in the low magnetic field limit ($\omega_c \tau \ll 1$)[3]. Experiments at higher fields ($\omega_c \tau > 1$) have been interpreted in terms of a coexistence of classical bulk conduction and quantum conduction in edge states [4-6], resulting in a breakdown of universal scaling of conductance fluctuations [7]. Fluctuations in high mobility samples near the minima of $R_{xx}$ in the quantum Hall regime have been explained in terms of inter edge state tunnelling via magnetic bound states [8,9]. However, in samples of mesoscopic size fluctuations from the bulk region can always be expected and an accurate separation of a possible contribution of the edge state region is difficult. An appropriate system for investigating the quantum Hall regime without the quantum Hall effect (QHE) are wide parabolic quantum wells (WPQW) with a sufficiently small subband splitting. In order to rule out UCF's from the bulk, also the Hall bars have to be of macroscopic size. The realization of WPQW's with a flat potential in the electron channel requires a parabolic bare potential which is difficult to obtain in hetero structures by MBE growth[10-12]. WPQW's can be made much easier by p-n-p structures according to the nipi-concept with PbTe [13,14]. An essential difference to "usual" quantum wells is that the subband splitting is much smaller than the Landau level (LL) - splitting even at relatively low magnetic fields. Consequently, even if the Fermi energy is already down in the lowest bulk - LL, still a number subbands can be occupied. PbTe is a narrow gap many valley semiconductor with four effective mass ellipsoids oriented parallel to the <111> directions with an anisotropy of about 10 [15]. For a magnetic field parallel to the <111> growth direction therefore we have two electron systems: one can lead to 2D-subbands with typically 1..2meV subband separation while the other one is still a 3D electron system [16,17].

## 2. Experiments:

The experiments have been performed in a super conducting magnet in fields up to 17 Tesla with a dilution refrigerator for temperatures down to 40mK. A standard lock-in technique was used (f = 13.7 Hz) and the constant ac-current was chosen to be $\leq$10nA.

The sample parameters are the following: width of the Hall bar w = 100μm, Hall bar length between voltage probes L = 200μm, electron mobility μ = 270000cm$^2$/Vs, bulk doping level in the electron channel $N_D$=3x10$^{17}$cm$^{-3}$ and effective width of the electron channel $d_n$ = 330 nm. Selective contacts to the embedded electron channel are made by evaporating Indium and subsequent diffusion [13].

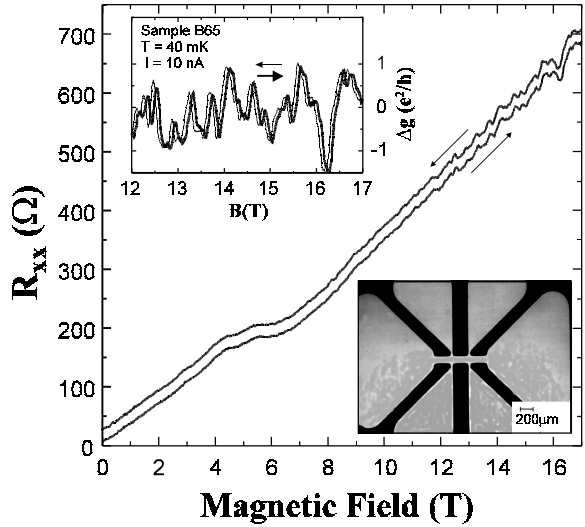

*Fig.1: typical Magnetotransport data, upper insert: Δg which was extracted from $R_{xx}$, lower insert: lateral sample structure*

Fig.1 shows data of a typical magneto transport experiment. $R_{xx}$ below B = 8 T exhibit Shubnikov - de Haas oscillations according to the bulk parameters of the electron channel [17]. For extracting the fluctuations only the upper field range was used. After obtaining ΔR, the CF have been calculated according to Δg = ΔR/R². Δg appears to be of the order of $e^2/h$.

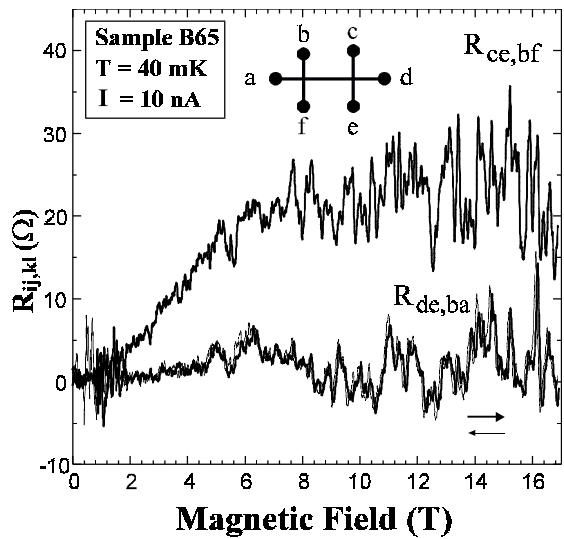

*Fig.2: non - local magnetotransport experiments for two different configurations*

Fig.2 shows the data of two different non-local configurations $R_{ij,kl}$. *ij* denotes the current contacts and *kl* the voltage probes. It is interesting to note that in configuration $R_{ce,bf}$ there is a direct path along the edge from contact c to b and from e to f. Following the sample edge in the second configuration $R_{de,ba}$, one can find always an unused contact between the current and the voltage contacts.

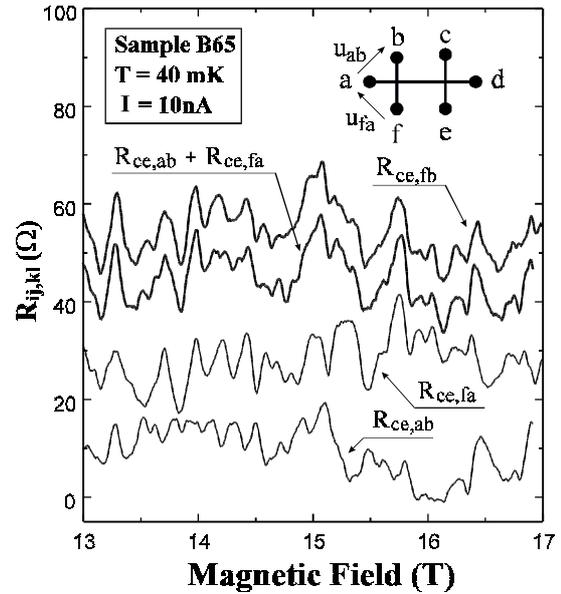

*Fig.3: non-local signals from different parts of the sample edge. The traces are shifted vertically for better visibility.*

Fig.3 shows data of configuration $R_{ce,bf}$ in which the fluctuations have been measured also between contacts f and a and contacts a and b separately. One can clearly see that $R_{ce,bf}$ is the sum of $R_{ce,fa}$ and $R_{ce,ab}$.

## 3. Discussion:

The increase of the non-local signal $R_{ce,bf}$ to an average value of about 20Ω with increasing field in Fig.2 indicates that edge channel conduction is established in the upper field range and also that a considerable amount of dissipation occurs between the contacts b and f. This increase does not appear in $R_{de,ba}$ although the amplitude of the fluctuations is nearly the same. The reason can be the presence of the unused contact between the current and voltage contacts, which provides the possibility for carriers in edge states to be scattered to the 3D electron system before they can reach the voltage probes. The individual patterns $R_{ce,fa}$ and $R_{ce,ab}$ of Fig.3 are considerably different. This strongly suggests that the fluctuations do not result from the bulk. However, also edge channel back scattering in the Hall bar should produce a significant similarity in the two patterns. The reason for the fluctuations is not yet clear but they have to be attributed definitely to the edge state region. Considering all these facts, the edge channels seem to enable phase coherent transport over the macroscopic distance between the contacts. It is important to note that this would not necessarily be in contradiction with the missing zeros in $R_{xx}$. The only requirement is the possibility for scattering of edge electrons to the 3D electron system at the contacts. In contrast, recently observed CF in other IV-VI-epilayers, particularly PbSe and $Pb_{1-x}Mn_xSe$ single layers of thickness ranging from 1.5 - 3 µm, exhibit a clearly non-universal amplitude which changes with the magnetic field [18,19]. However, there are no non-local data which would allow a more detailed comparison with our results.

## 4. Summary

We have performed magneto transport experiments with Hall bars of macroscopic size. Conductance fluctuations are observed which are close to $e^2/h$. With non-local experiments we have demonstrated that edge channel transport plays an important role and that the observed fluctuations have to be attributed to edge channel effects.

## 5. Acknowledgements

Financial support by Fonds zur Förderung der wissenschaftlichen Forschung (FWF) Vienna (Project: P10510 NAW) and Steiermärkischer Wissenschafts und Forschungslandesfonds Graz.